# Electronic Stopping of Slow Protons in Transition and Rare Earth Metals: Breakdown of the Free Electron Gas Concept

D. Roth[1], B. Bruckner[1], M.V. Moro[1,7], S. Gruber[1], D. Goebl[1], J.I. Juaristi[2,3,4], M. Alducin[2,3], R. Steinberger[5], J. Duchoslav[5], D. Primetzhofer[6], and P. Bauer[1,2]

[1]Johannes-Kepler Universität Linz, IEP-AOP, Altenbergerstraße 69, A-4040 Linz, Austria

[2]Donostia International Physics Center DIPC, P. Manuel de Lardizabal 4, 20018 San Sebastián, Spain

[3]Centro de Física de Materiales CFM/MPC (CSIC-UPV/EHU), P. Manuel de Lardizabal 5, 20018 Donostia-San Sebastián, Spain

[4]Departamento de Física de Materiales, Facultad de Químicas, Universidad del País Vasco (UPV/EHU), Apartado 1072, 20018 San Sebastián, Spain

[5]Christian Doppler Laboratory for Microscopic and Spectroscopic Material Characterization, Zentrum für Oberflächen- und Nanoanalytik (ZONA), Johannes-Kepler Universität Linz, Altenbergerstraße 69, A-4040 Linz Austria

[6]Institutionen för Fysik och Astronomi, Uppsala Universitet, Box 516, S-751 20 Uppsala, Sweden

[7]Instituto de Física, Universidade de São Paulo, Rua do Matão 1371, 05508-090, São Paulo, Brasil

**Abstract**

The electronic stopping cross sections (SCS) of Ta and Gd for slow protons have been investigated experimentally. The data are compared to the results for Pt and Au, to learn how electronic stopping in transition and rare earth metals correlates with features of the electronic band structures. The extraordinarily high SCS observed for protons in Ta and Gd cannot be understood in terms of a free electron gas model, but are related to the high densities of both occupied and unoccupied electronic states in these metals.

When ions propagate in matter, they are decelerated due to interaction with both, nuclei and electrons – in other words, by nuclear and electronic stopping, respectively. The mean energy loss per path length is given by the deceleration force, i.e., the stopping power $S = dE/dx$. To describe the interaction with atoms or molecules, often the stopping cross section (SCS) $\varepsilon = 1/n \cdot dE/dx$ is used, where $n$ represents the atomic or molecular density of the target material, respectively. Precise knowledge of electronic stopping is important in many different fields like outer space exploration (space weathering), nanotechnology (ion beam patterning), fusion research (plasma-wall interactions), or medicine (radiation therapy) [1, 2, 3, 4]. Thus, the SCS is a quite fundamental quantity, for which it is mandatory to have a detailed understanding of the prevailing energy loss mechanisms.

At high ion velocities $v \gg v_F$ ($v_F$ denotes the Fermi velocity of the target electrons), energy loss of ions is dominated by electronic stopping. In this regime, light ions represent only a weak perturbation for the target electrons, and first order theoretical models are sufficient for a precise description of the energy loss process [5, 6, 7, 8]. For low ion velocities $v \leq v_F$, both electronic and nuclear collisions contribute to the stopping power. In this regime, the electronic energy loss is predominantly due to interaction with valence electrons. For a free electron gas (FEG), the stopping power is proportional to the ion velocity, $S = Q(Z_1, r_s) \cdot v$, where the friction coefficient $Q$ depends on the atomic number of the ion $Z_1$ and the Wigner-Seitz radius of the FEG $r_s = (3/4\pi n_e)^{1/3}$, with the FEG density $n_e = N_{val} \cdot n$, where $N_{val}$ stands for the (effective) number of valence electrons per atom [9]. At low velocities, even for light ions, the electronic interaction is so strong that nonlinear theoretical models have to be applied to solve this complex many-body problem. In their seminal work, Echenique *et al.* found the quantum mechanically correct solution for the stopping power of static ions in a FEG [10]. When Mann *et al.* showed that for many target materials the existing experimental data could be well reproduced for ion velocities up to $v_F$ by this nonlinear model, the physics of slow ion stopping was seemingly well understood [11].

Limitations of the applicability of the FEG model have been discovered for noble metals at very low proton velocities $v \ll v_F$: deviations from velocity proportionality of $\varepsilon$, e.g., for H[+] in Au at $v \geq 0.2$ a.u. [12] were attributed to the finite excitation threshold of the $d$ band (for Au, 2 eV [13, 14]). Recently, time-dependent density functional (TD-DFT) calculations of electronic stopping of protons in Au in channelling geometry confirmed this interpretation and revealed a low but finite contribution of the $d$ band to electronic stopping down to $v = 0.15$ a.u. [15]. Still today, even for protons traversing in a real metal (beyond a FEG) it is an open question which are the processes dominant in the stopping power at low velocities [16].

In fact, for noble metals, there exist two distinct regimes where the non-linear FEG model [10] can be applied by assuming a proper FEG density. For Au, at proton velocities $v < 0.2$ a.u., electronic stopping is predominantly due to excitation of the $6s^1$ electron, which can be described as a low density FEG ($r_s = 3.01$ a.u.), whereas at $v \approx v_F$, all conduction electrons are excited as a FEG of effective density as deduced from the plasmon frequency ($r_{s,eff} = 1.49$ a.u.) [17, 18]. Metals like Pt feature a high density of states (DOS) up to the Fermi energy $E_F$ due to $d$ electrons, followed by a low density of free states at $E > E_F$. For Pt, the use of $r_{s,eff} = 1.63$ a.u. ($N_{val,eff} = 5.6$) leads to good agreement between experimental and FEG stopping power at all velocities up to $v_F$ [17, 19].

In this Letter, we present evidence that the FEG model fails when applied to metals with $d$ or $f$ bands of high DOS below and above $E_F$, like in transition and rare earth metals. To this aim, we present electronic SCS data for low-velocity protons in polycrystalline Ta ([Xe]$4f^{14}5d^36s^2$) and Gd ([Xe]$4f^75d^16s^2$); for Gd, the velocity range was extended to cover also the stopping maximum.

Since transition and, especially, rare earth metals are chemically reactive, information on impurities and their chemical states is essential. To this aim, purity of both samples (250 μm 99.99 % foils of Ta and Gd, respectively) was analyzed, by time-of-flight elastic recoil detection (TOF-ERD) and Rutherford backscattering spectrometry (RBS) at Uppsala University, as well as by x-ray photoelectron spectroscopy (XPS) and Auger electron spectroscopy (AES) in ZONA (JKU). For Ta, XPS and TOF-ERD showed no significant bulk concentrations of light impurities. Also the Gd foil was found to be a pure metal covered by a ~ 1 µm thick nearly stoichiometric oxide layer. After removing the $Gd_2O_3$ layer, very fast oxygen uptake was observed even at UHV conditions (base pressure: $5 \times 10^{-10}$ mbar), in accordance with [20, 21]. XPS indicated that also with adsorbed oxygen on top Gd is present in metallic form. The energy loss measurements were performed at the IEP in Linz employing the UHV time-of-flight low energy ion scattering (TOF-LEIS) setup ACOLISSA [22] and the RBS setup of the AN 700 accelerator [23]. In ACOLISSA, both samples were cleaned by means of 3 keV $Ar^+$ sputtering in order to remove adsorbed hydrocarbons and the native oxide layer. Prior to transfer into vacuum, the Gd sample was mechanically polished in Ar atmosphere (grain size ~ 5 µm). Surface composition was checked by AES. Sputtering cycles were repeated until the Auger intensities corresponding to O, Gd, and$$ Ta stayed constant (95 % Ta, 5 % O, and 82 % Gd, 18 % O, respectively).

TOF-LEIS spectra of Gd, Ta, and Au were acquired by use of beams of hydrogen and deuterium ions (monomers and dimers) in the range of 0.5 keV/u – 10 keV/u (scattering angle $\theta = 129°$) at a base pressure of $< 4 \times 10^{-11}$ mbar. Even for Gd, no change in the LEIS spectra was observed when comparing the first and the last measurement within one data collection cycle. To cover the Bragg peak of $\varepsilon_{Gd}$, RBS spectra of Gd and Au were recorded employing atomic beams of $H^+$ and $D^+$ in the range 30 keV/u – 600 keV/u ($\theta = 154.6°$).

The SCS of the materials of interest ("$X$") was deduced from the energy spectra for projectiles backscattered from $X$ and from Au as a reference sample ("ref"). This approach makes use of the fact that the ratio of the spectrum heights, $N_X/N_{ref}$, contains information on $[\varepsilon]_{ref}/[\varepsilon]_X$ [24]. Because of the influence of multiple scattering at low energies, the ratios of the experimental spectra in an energy interval close to the kinematic onset [25] were compared to the results from corresponding Monte Carlo simulations (TRBS, [26]). In the simulations, the ZBL potential [27] was used, and $\varepsilon_{Au}$ was set to match the experimental value (SCS data from [14]), leaving $\varepsilon_X$ as the only parameter, which was optimized until convergence was reached; i.e., the heights of the resulting spectra changed by less than 3 %.

Since RBS spectra of the polished Gd sample showed an oxygen enriched surface layer of considerable thickness [28], the LEIS data were evaluated applying Bragg's rule [29] taking the SCS data of oxygen from SRIM-2013 [30]. This correction results in a lowering of $\varepsilon_{Gd}$ by 8 %. In case of the RBS data, energy windows in the evaluation were chosen such that only projectiles backscattered in the bulk Gd were considered. For E > 200 keV, spectra were simulated by use of the program SIMNRA [31] employing a screened Coulomb potential [32] and the dual scattering model.

In Fig. 1, experimental and simulated spectra of 10 keV $D^+$ scattered from Au and Gd are shown, exhibiting excellent agreement between experiment (open symbols) and simulation (solid lines) due to optimization of $\varepsilon_{Gd}$. Note that the simulated spectra were convoluted with a Gaussian to account for the experimental resolution. Dashed lines represent the evaluated energy window. To demonstrate the sensitivity of the plateau height of Gd to $\varepsilon_{Gd}$, two additional simulations are shown, where $\varepsilon_{Gd}$ was varied by ± 20 % (short-dashed spectra).

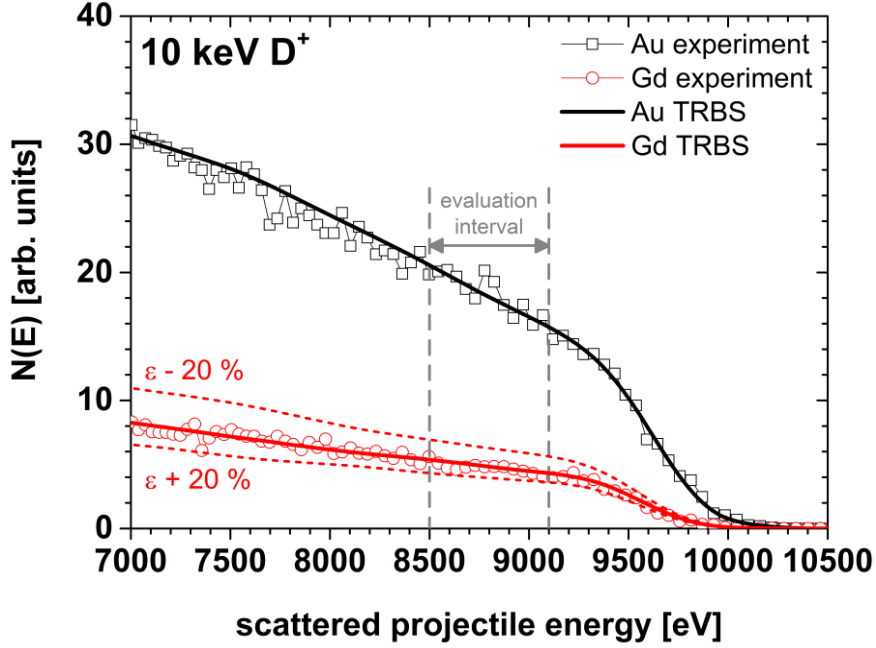

*FIG. 1 Experimental and simulated energy spectra of 10 keV $D^+$ scattered from Au and Gd are shown. TRBS simulations (solid lines) and experimental spectra (open symbols) coincide when experimentally deduced SCS are employed. Dashed lines represent the energy interval used in the evaluation, short-dashed spectra refer to simulations in which $\varepsilon_{Gd}$ was varied by ±20 %.*

The electronic SCS of Gd for H ions (protons and deuterons) is shown as function of the proton energy in Fig. 2. LEIS and RBS data are very well consistent. No isotope or vicinage effects were observed; thus, we refer in the following discussion for the projectiles as to protons. For Gd, the maximum SCS is found to be very high ($\varepsilon_{Gd} \sim 48 \times 10^{-15}$ eV cm$^2$/atom), while the position of the stopping maximum is rather low ($E_{max} \sim 80$ keV). At higher energies, the RBS data coincide with data from literature [33, 34, 35, 36]. The bold solid line is a fit using the semiempirical model described in [27]. The overall consistency of the results and the agreement with literature gives confidence in the experiment and the evaluation procedure, even for a chemically reactive metal like Gd.

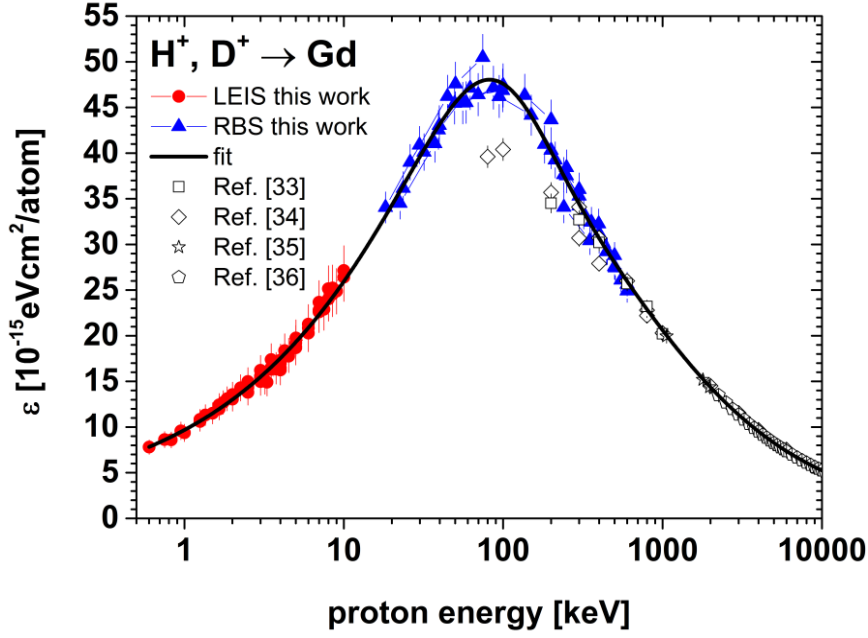

*FIG. 2 Electronic SCS of Gd for H ions (protons and deuterons) are shown as function of the proton energy in the range 0.5 keV – 10 MeV (LEIS data: full circles, RBS data: full triangles). Also displayed is a fit to the complete data set. Open symbols refer to data from Refs. [33 – 36].*

In Fig. 3, the low-velocity SCS of Gd, Ta, Pt, and Au for protons are presented as function of the ion velocity. The SCS of both, Gd and Ta are considerably higher than that of Pt and Au: at $v = 0.2$ a.u., $\varepsilon_{Gd}$ and $\varepsilon_{Ta}$ exceed $\varepsilon_{Au}$ by a factor of 3.8 and 2.6, respectively. The statistical uncertainty of the SCS in the LEIS regime is estimated to be < 10 % per data point. Systematic errors are estimated to be < 10 % and mainly due to the correction for impurities. The statistical uncertainty of the present RBS data is 5 % per data point; systematic errors due to the correction for impurities are < 5 %.

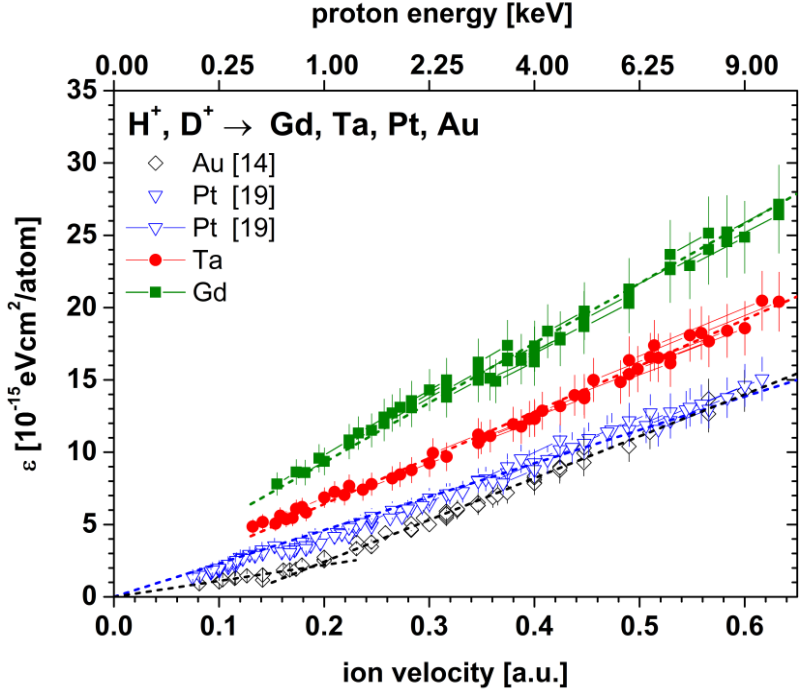

*FIG. 3 Electronic SCS of Au, Pt, Ta, and Gd for protons are displayed as function of the ion velocity in atomic units. The upper abscissa denotes the corresponding proton energies.*

While at $v < 0.2$ a.u., the nonlinear FEG model [10] successfully describes the stopping ratio $\varepsilon_{Pt}/\varepsilon_{Au} \approx 1.7$ by use of effective FEG densities [17], for Gd and Ta, the FEG concept fails: by use of $Q_H(r_s)$ from theory [10, 37], the slopes of the experimental data can be converted to $r_{s,expt}$ values of 1.14 and 2.03 for Ta and Gd, respectively. These low $r_{s,expt}$ values correspond to huge values of $N_{val,expt}$. For Ta, $N_{val,expt} = 19.6$, which is a figure without any physical meaning. From the electronic configuration of Ta, $[Xe]4f^{14}5d^{3}6s^{2}$, one should expect $N_{val} = 5$, since the $f$ electrons are located at ~ 26 eV below $E_F$ [38] and do neither contribute to low energy electron-hole (e-h) pair excitations nor to the plasmon oscillation. The measured plasmon energy ($\hbar\omega_{p,Ta} = 20.8$ eV [39]) corresponds to $r_{s,eff} = 1.72$ ($N_{val,eff} = 5.7$), which is consistent with this expectation, in contrast to $N_{val,expt} = 19.6$ as deduced from the SCS.

For Gd, the value for $r_{s,\text{expt}}$ is almost identical to that of Al [40]. Hence, $S_{\text{Gd}} \cong S_{\text{Al}}$, in accordance with the fact that $N_{\text{val}} = 3$ for both Al and Gd. However, the atomic density $n_{\text{Gd}}$ is lower than $n_{\text{Al}}$ by a factor of 2. Consequently, $r_{s,\text{expt}} = 2.03$ corresponds to $N_{\text{val,expt}} = 6.4$ for Gd – in clear contradiction to the FEG model, on condition that the seven $f$ electrons do not participate in the electronic interactions of slow protons (neither in the e-h pair excitation nor in the plasmon excitation). Note that the experimental plasmon energy of Gd ($\hbar\omega_{p,\text{Gd}} = 14$ eV [41]) corresponds to $r_{s,\text{eff}} = 2.24$ ($N_{\text{val,eff}} = 4.7$).

Since the observed features in electronic stopping should be related to electronic properties of the materials, the electronic DOS of the investigated metals were calculated by means of DFT, employing the VASP code [42, 43]. For Au, Pt, and Ta, the PBE exchange-correlation functional has been used [44]. In the case of Gd, a PBE + U approach was used with the recommended values U = 6.7 eV and J = 0.7 eV [45, 46] in order to assure an adequate description of the 4$f$ orbitals. In all cases, the core-valence electrons interaction is treated by the projected augmented wave (PAW) potentials [47, 48]. The energy cutoffs for the plane wave basis sets are 287.4 eV for Au, 287.9 eV for Pt, 279.6 eV for Ta, and 320.6 eV for Gd. The Brillouin zone is sampled by a 17 × 17 × 17 Monkhorst-Pack grid of **k** points [49] for Au, Ta, and Pt, and a 17 × 17 × 11 grid for Gd. In evaluating the DOS, the occupancies of the electronic states are determined with the tetrahedron method.

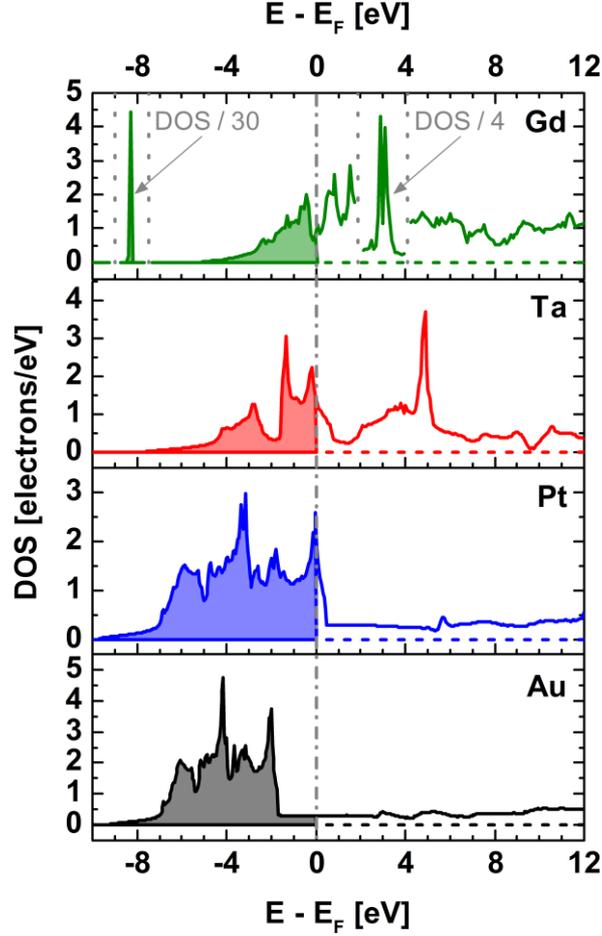

*FIG. 4 Electronic densities of states (DOS) are shown for the conduction bands of Au, Pt, Ta, and Gd, as function of E - $E_F$. The filled part of the DOS corresponds to the occupied states. For the DOS of Gd dotted lines indicate below and above $E_F$ the energy intervals, in which the high features due to the f band were scaled down by factors of 30 and 4, respectively.*

In Fig. 4, the electronic DOS of Au, Pt, Ta, and Gd are displayed as a function of $E - E_F$. Amongst these metals, only for Au, the DOS at $E \approx E_F$ is dominated by the *s* electrons, while for the other metals, the density at $E < E_F$ is considerably higher, due to the contribution of the *d* bands. In our calculation, the occupied *f* states of Gd are located at ~ 8.3 eV below $E_F$. From this, one may assume that at $v \leq 0.2$ a.u., the *f* electrons do not participate in electronic stopping, similarly as the *d* band in Zn is not excited at these low velocities [17]. On the other hand, an increase of the slope of $\varepsilon_{Gd}$ is not observed at $v > 0.2$ a.u. (as for Zn). Thus, excitation of the *f* band of Gd is either possible already at very low velocities or becomes considerable at higher velocities only. The conduction band of Gd has parabolic shape similarly as for FEG like metals, e.g., Al, but the DOS ($E_F$) of Gd exceeds that of Al by a factor of 4.5, while the occupied part of the band is much narrower. At ~ 3 eV above $E_F$, the DOS exhibits a prominent peak corresponding to seven electrons due to the *f* band, which may increase the efficiency of the e-h pair excitation at very low ion velocities (e.g., $v \approx 0.1$ a.u.) even further. This may cause a steeper slope of the Gd SCS data at $v < 0.25$ a.u..

As a next step, we compare the DOS of occupied and unoccupied states of the metals presented in Fig. 4. The DOS integrated from − 10 eV up to $E_F$ yield values in the range from 5 electrons (Ta) to 11 electrons (Au), as expected from $N_{val}$. Above the Fermi level, the DOS is low for Pt and Au (*s*- and *p*-states only), while Ta and Gd exhibit a high *d*-electron density; the DOS integrated from $E_F$ up to 10 eV ranges from 3.3 electrons (Au, Pt) and 6.9 electrons for Ta to ~ 17 electrons (Gd). Note that for Ta, the DOS has nothing in common with a FEG like parabola, neither below nor above $E_F$. Given these facts, our stopping data can be qualitatively understood in the following way: on the one hand, Gd and Ta exhibit high DOS below and above $E_F$, thereby allowing for low energy excitations with high probabilities. On the other hand, for Pt and Au, the low DOS above $E_F$ will be responsible for the low SCS, as compared to Ta or Gd. Note that this argument should hold true for the other transition and rare earth metals too, as long as both, DOS ($E < E_F$) and DOS ($E > E_F$) are high. In fact, also at the stopping maximum, the combination of high DOS below and above $E_F$ helps to explain the very high SCS of Gd (excitation of *f* electrons).

To conclude, electronic stopping of slow protons in transition and rare earth metals is so efficient that any FEG model fails to describe the slowing down process, ending up in absurdly high effective FEG densities, or in other words, in an unphysical number of valence electrons. For instance, both the rare earth metal Gd and aluminum exhibit $N_{val} = 3$ and virtually the same FEG density parameter ($r_{s,expt} \approx 2.03$) deduced from the measured SCS, but for Gd, an effective number of 6.4 valence electrons results due to its low atomic density. It is interesting to note that the DOS at $E_F$ is much higher for Gd and Ta compared to Au and Pt. For the latter, the low DOS above $E_F$ may be responsible for the much lower values of $N_{val,eff}$ (for Pt, 5.6 instead of 10). Therefore, a FEG description of electronic stopping collapses when applied to transition and rare earth metals featuring a high DOS (below *and* above $E_F$). Elaborate many-body theoretical models like TD-DFT are required to thoroughly understand the underlying physical mechanisms and to find possibly a way of predicting realistic stopping power values for rare earth and transition metals even for simple ions.

Financial support of this work by the Austrian Science Fund (FWF-Project No. P22587-N20 and FWF-Project No. P25704-N20) is gratefully acknowledged. M.V.M. acknowledges financial support from São Paulo/Brazil funding agency FAPESP under Contract No. FAPESP2013/09105-0. A research infrastructure fellowship of the Swedish Foundation for Strategic Research (SSF) under Contract No. RIF14-0053 supporting accelerator operation is acknowledged. P.B. expresses his gratitude for the kind hospitality at the DIPC in San Sebastián. We are grateful to Pedro Echenique, Andres Arnau and Peter Zeppenfeld for their interest in this work and for inspiring discussions.

**References and remarks:**